\documentclass{aa}
\usepackage{txfonts}
\usepackage{natbib}
\bibpunct{(}{)}{;}{a}{}{,} 

\usepackage{graphicx}

\newcommand{\cxo}{{\it Chandra}}

\begin{document}
   \title{X-rays from \object{HH\,210} in the Orion Nebula}

\titlerunning{X-rays from HH\,210 in the Orion nebula}
\authorrunning{Grosso et al.}


   \author{N.\ Grosso\inst{1}
      \and E.~D.\ Feigelson\inst{2} 
      \and K.~V.\ Getman\inst{2}
      \and J.~H.\ Kastner\inst{3}
      \and J.\ Bally\inst{4}
      \and M.~J.~McCaughrean\inst{5,6}
        }


   \institute{
              Laboratoire d'Astrophysique de Grenoble, 
              Universit{\'e} Joseph-Fourier,
              F-38041 Grenoble cedex 9, France\\
              e-mail: {\tt Nicolas.Grosso@obs.ujf-grenoble.fr}
         \and Department of Astronomy and Astrophysics, 
              Pennsylvania State University, 
              University Park, PA~16802, USA
         \and Center for Imaging Science, Rochester Institute of Technology, Rochester, 
              New York 14623-5604, USA
         \and Center for Astrophysics and Space Astronomy, 
              University of Colorado at Boulder, CB 389, 
              Boulder, CO 80309, USA
         \and School of Physics, University of Exeter, Stocker Road,
              Exeter EX4 4QL, Devon, UK
         \and Astrophysikalisches Institut Potsdam,
              An der Sternwarte 16,
              D-14482 Potsdam, Germany
            }

   \date{}

   \abstract{We report the detection during the {\it Chandra Orion Ultradeep
Project} (COUP) of two soft, constant, and faint X-ray sources associated
with the Herbig-Haro object HH\,210. HH\,210 is located at the tip of
the NNE finger of the emission line system bursting out of the
BN-KL complex, northwest of the Trapezium cluster in the OMC-1
molecular cloud. Using a recent H$\alpha$ image obtained with the {\it
ACS} imager on board {\it HST}, and taking into account the known
proper motions of HH\,210 emission knots, we show that the position of
the brightest X-ray source, \object{COUP\,703}, coincides with the
emission knot 154-040a of HH\,210, which is the emission knot of
HH\,210 having the highest tangential velocity
(425\,km\,s$^{-1}$). The second X-ray source, \object{COUP\,704}, is
located on the complicated emission tail of HH\,210 close to an
emission line filament and has no obvious optical/infrared
counterpart. Spectral fitting indicates for both sources a plasma
temperature of $\sim$0.8\,MK and absorption-corrected X-ray
luminosities of about $10^{30}$\,erg\,s$^{-1}$ (0.5--2.0\,keV). These
X-ray sources are well explained by a model invoking a fast-moving,
radiative bow shock in a neutral medium with a density of
$\sim$12000\,cm$^{-3}$. The X-ray detection of COUP\,704 therefore
reveals, in the complicated HH\,210 region, an energetic shock not yet
identified at other wavelengths.
               }

\maketitle

\section{Introduction}

\begin{figure}[!ht]
\centering
\includegraphics[width=\columnwidth]{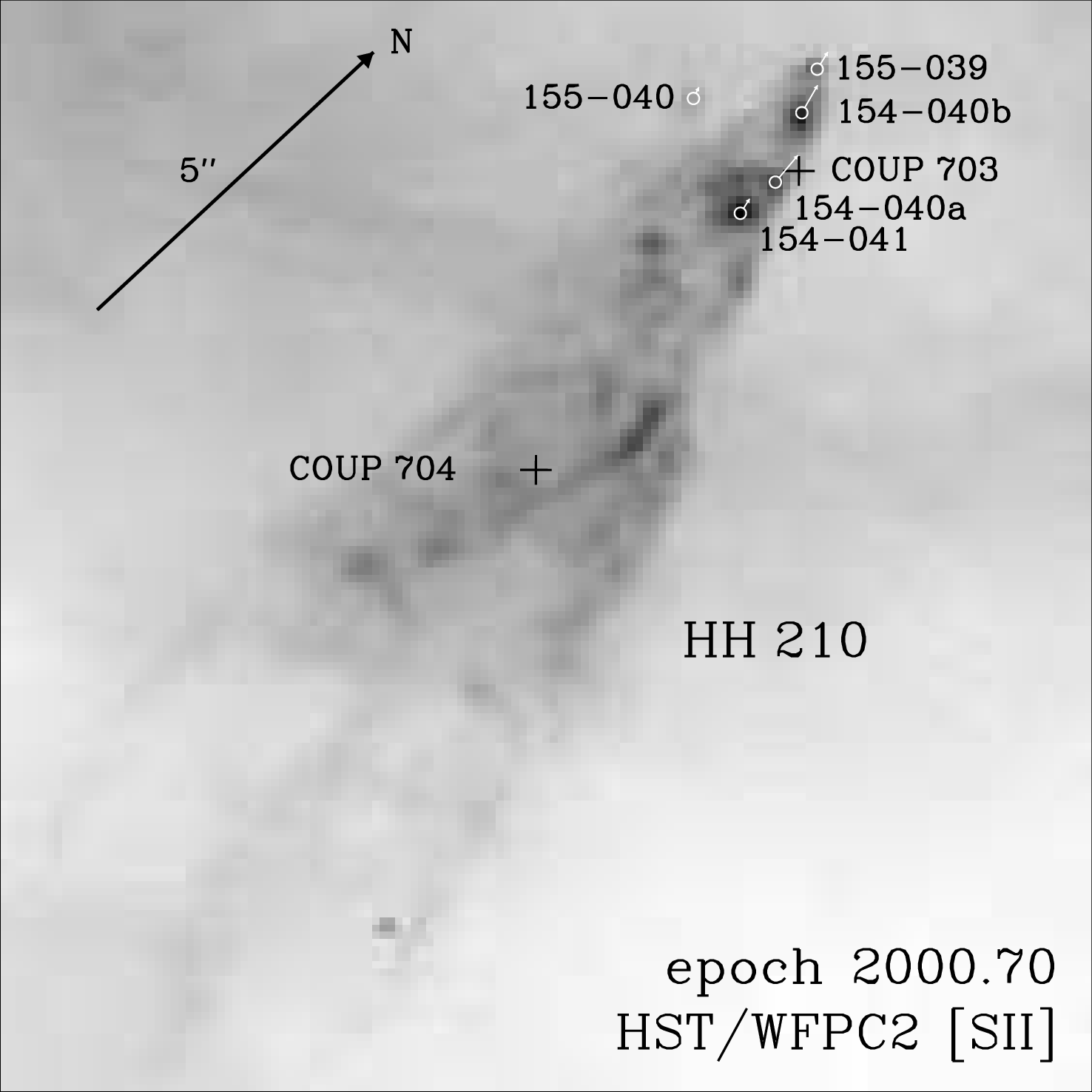}
 \caption{{\it HST/WFPC2} image of HH\,210 for epoch 2000.70 in
[\ion{S}{ii}] \citep[adapted from Fig.~2 of][]{doi02}. Each pixel is
$0\farcs1$ on a side, the color stretch is linear. HH\,210 shows a
typical bow shock structure containing numerous emission knots and
filaments. The two crosses mark epoch 2003.04 positions of COUP\,703
and COUP\,704 \citep{getman05b}; cross sizes indicate the total positional
uncertainties of these X-ray sources ($0\farcs14$). The white
arrows indicate velocity and epoch 2003.04 position for emission knots
with proper-motion measured in [\ion{S}{ii}] by \citet{doi02}. The
white $0\farcs08$-radius circles indicate the optical resolution of
{\it HST}.
}
\label{map}
\end{figure}

Outflow activity is ubiquitous in young stellar objects (YSO), and
intimately connected to the accretion process building up the mass of
stars. The interaction of outflow from YSO and the interstellar medium
produces bow shocks associated with optically luminous small nebulae,
called Herbig-Haro (HH) objects (\citealt{herbig50,haro52};\citealp[
see review by][]{reipurth01}). Optical emission from HH objects
includes Hydrogen Balmer lines, [\ion{O}{i}], [\ion{S}{ii}],
[\ion{N}{ii}], [\ion{O}{iii}], tracing plasma with temperatures of
several $10^3$ to $10^5$\,K. During these last years a handful of HH
objects have been detected in X-rays, tracing plasma with temperatures
of a few $10^6$\,K, as expected from the high velocities of these
shocks \citep[e.g.,][]{raga02}. The emission knot H of \object{HH\,2}
was the first detected in X-rays; it also exhibits strong H$\alpha$
and free-free radio continuum emission
\citep{pravdo01}. \object{HH\,80} and \object{HH\,81} are excited by a
massive protostar, and are among the most luminous HH objects in the
optical; emission knots~A and G/H of HH\,80, and emission knot~A of
HH\,81 are now the most luminous known in X-rays \citep{pravdo04}. 

The soft X-ray emission from L1551~IRS5, associated with
\object{HH\,154} by \citet{favata02}, is now considered as arising
from a shock at the base of the jet of this protostar binary
\citep{bally03,bonito04}. Recently, soft X-ray excesses were found in
the X-ray spectra of the \object{`Beehive' proplyd} \citep{kastner05}
and \object{DG\,Tau} \citep{guedel05}, two YSOs with jets. 
Remarkably, in both sources, this soft and constant X-ray emission 
suffers less extinction than the harder and variable X-ray component
likely emitted by the stellar corona, suggesting that it comes from
shocks along the base of these jets. 

In January 2003, the {\it Chandra Orion Ultradeep Project} (COUP) was
carried out with ACIS-I \citep{garmire03} on board the \cxo{} X-ray
Observatory \citep{weisskopf02}. COUP consisted of a nearly continuous
exposure over 13.2\,days centered on the Trapezium cluster in the
Orion Nebula, yielding a total on-source exposure time of 838\,ks,
i.e.\ 9.7\,days \citep{getman05b}. Among the 1616 COUP X-ray sources,
there are only about fifty X-ray sources with the median energy of
their X-ray photons lower than 1\,keV, i.e.\ having a very soft X-ray
spectrum. The bulk of this soft X-ray source sample is associated with
optical/infrared stars; two sources may be spurious; and two other
sources without counterparts may be newly discovered very low mass
members of the Orion nebula cluster. Of the very soft
sources, only COUP\,703 and COUP\,704 are possibly associated with an
HH object, HH\,210 \citep{getman05a,kastner05}. More generally there
is in COUP no other X-ray source associated with HH objects
\citep{getman05a}. 

HH\,210 is located 2.7\arcmin-north of $\theta^1$\,Ori\,C in the OMC-1
molecular cloud, and is one of the brightest HH objects in this area
revealed by emission of [\ion{O}{i}] \citep{axon84,taylor86}. It
belongs to the powerful CO outflow from the BN-KL region, which
produces a spectacular set of H$_2$ bow shocks and trailed wakes; it
is located at the tip of the NNE finger which is one of the
brightest emission systems in [\ion{Fe}{ii}] and lacks a trailed H$_2$
wake \citep{allen93}. HH\,210 has a pronounced bow shape containing
numerous knots and complicated filaments in [\ion{S}{ii}] (see
Fig.~\ref{map}).

We present in Sect.~\ref{observation} the astrometry and spectral analysis
of the X-ray sources associated with HH\,210; we discuss in
Sect.~\ref{discussion} the origin of these X-ray emissions.

\section{Astrometry and spectral analysis of the X-ray sources
associated with HH\,210}
\label{observation}

\begin{figure}
\centering
\includegraphics[width=\columnwidth]{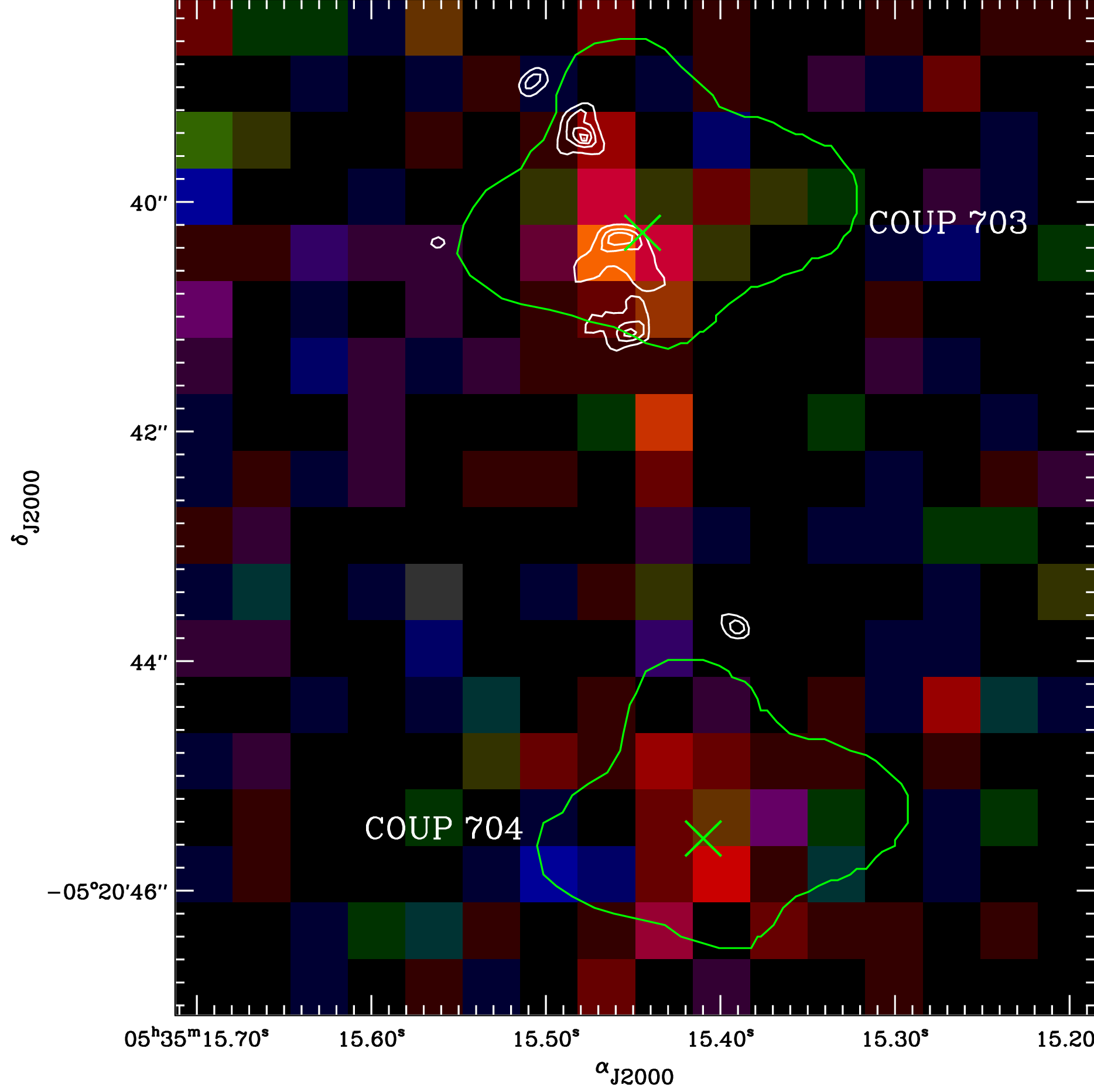}
 \caption{COUP view of HH\,210. Red, green, and blue represent photons
in the 0.5--1.7\,keV, 1.7--2.8\,keV, and 2.8--8.0\,keV energy bands,
respectively \citep{lupton04}. Green crosses and countours mark the
centre of COUP sources and source extraction polygons ($\sim$87\%
encircled energy using PSF at 1.5\,keV). Cross sizes indicate the
total positional uncertainties of these X-ray sources
($0\farcs15$). White contours show H$\alpha$+[\ion{N}{ii}] knots
observed with {\it HST/ACS} \citep[$0\farcs05$-pixels on a side; epoch
2004.05,][]{bally05}, which were moved individually to epoch 2003.04
using proper-motions of \citet{doi02}. COUP\,703 coincides with
154-040a~HH\,210, the emission knot having the highest tangential
velocity (425\,km\,s$^{-1}$; see Fig.~\ref{map}).
}
\label{trichro}
\end{figure}

\begin{figure*}[!ht]
\begin{tabular}{@{}cc@{}}
\includegraphics[width=12cm]{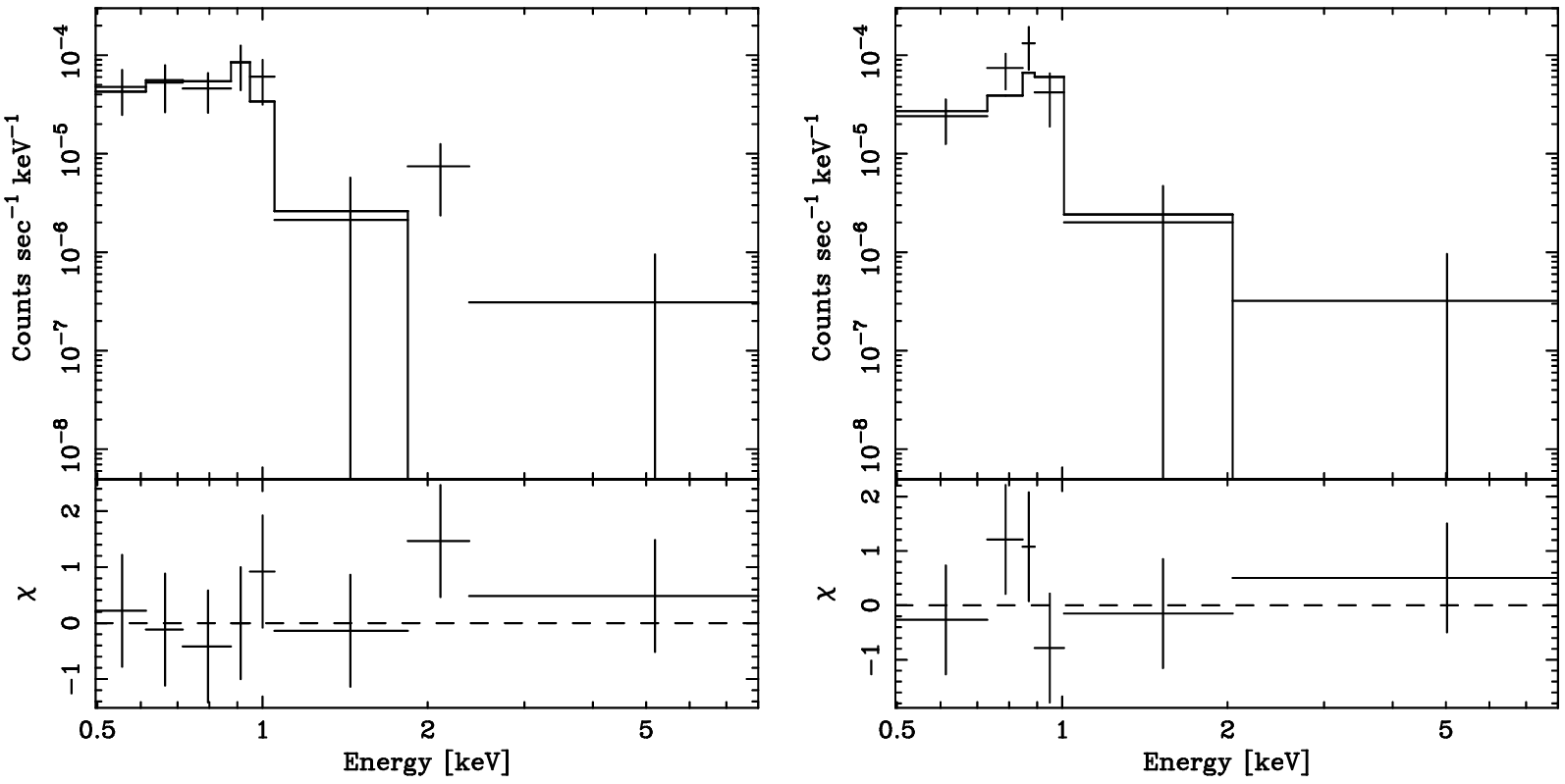} & \includegraphics[width=5.5cm]{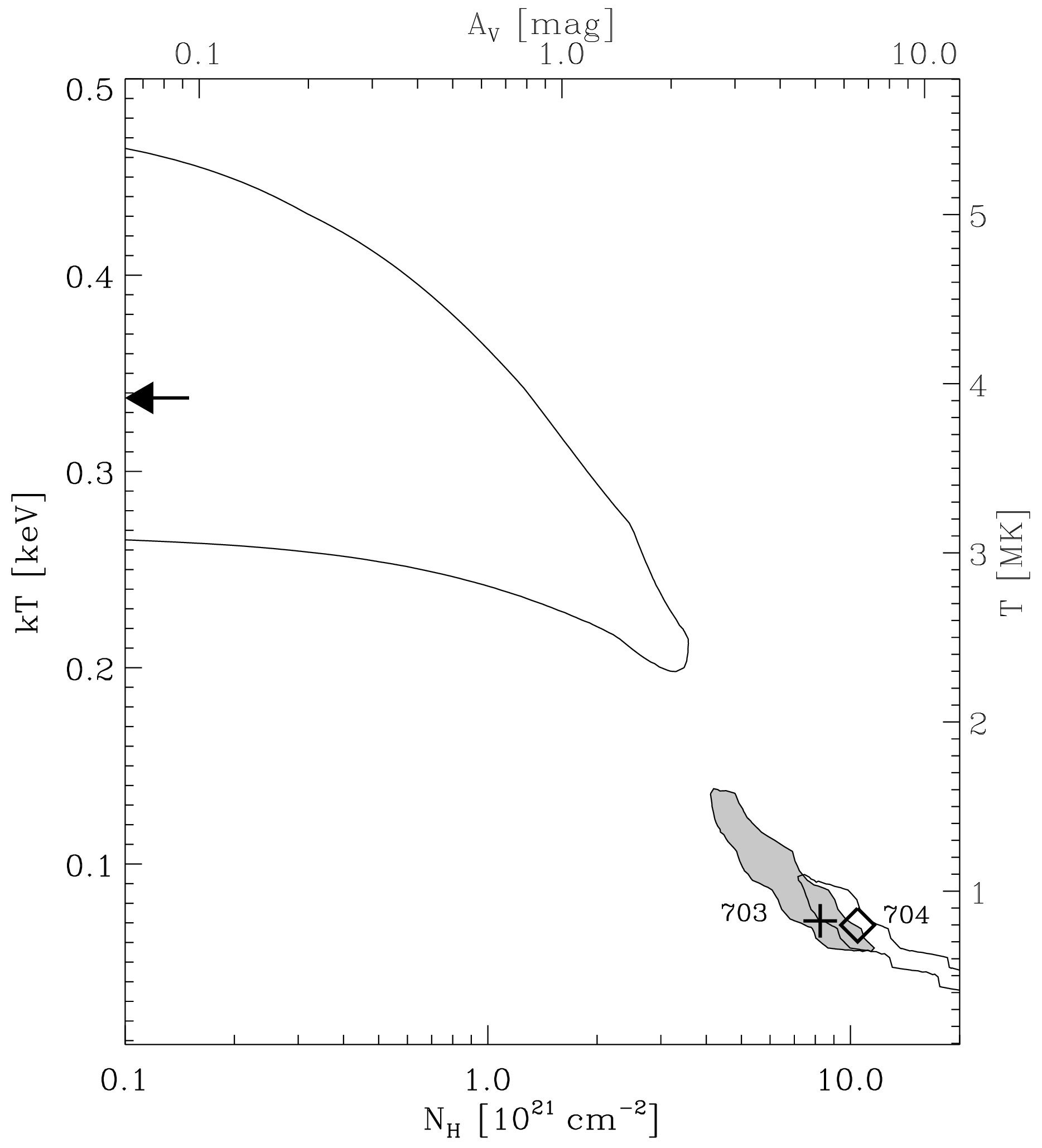}
\end{tabular}
\caption{X-ray spectra of COUP\,703 (left) and COUP\,704 (middle),
and confidence regions of the plasma parameters (right). Each
spectral bin contains at least 5 events. We used for the spectral fitting
a collisionally-ionized plasma model ({\tt APEC}),
assuming a plasma with solar elemental abundance \citep{anders89}, 
with X-ray absorption model \citep[{\tt WABS};][]{morrison83}. 
Fit parameters are given in Table~\ref{fit}. In the right panel,
the grey region indicates for COUP\,703 the confidence regions at the
68\% level of the plasma temperature and the absorption. The plus sign
and diamond mark the best parameter values for COUP\,703 and COUP\,704
(high $N_{\rm H}$/low $kT$ solution), respectively. The arrow
indicates the high temperature value for the null absorption solution
of COUP\,704.
}
\label{spectra}
\end{figure*}

\begin{table*}[!ht]
\caption{Spectral parameters of X-ray sources associated with HH\,210
using a {\tt WABS$\times$APEC} plasma model with solar elemental
abundance. We used Chi statistics with
standard weighting. Confidence ranges at the 68\% level
($\Delta\chi^2=1$; corresponding to $\sigma=1$ for Gaussian
statistics) are given in parentheses. $Q$ in Col.~(8) is the
probability that the best-fit model matches the data, given the value
of $\chi^2$ and $\nu$, the degree of freedom in Col.~(7). The emission
measure in Col.~(6) and the logarithm of the X-ray luminosities in the
0.5--2\,keV energy range observed/corrected for absorption in
Col.~(9) were computed assuming a distance of 450\,pc.}
\label{fit}
\vspace{-0.25cm}
\centering
\begin{tabular}{ccccccccc}
\hline\hline
\smallskip
COUP & CXOONCJ & Net & $N_{\rm H}$ & $kT$ & $EM$ & $\chi^2_\nu$ ($\nu$) & $Q$ & $\log L_{\rm S}/\log L_{\rm S,c}$ \\
\#   &         & conts & $10^{21}$\,cm$^{-2}$ & keV & $10^{55}$\,cm$^{-3}$ & & \% & erg\,s$^{-1}$ \\
(1) & (2) & (3) & (4) & (5) & (6) & (7) & (8) & (9) \\
\hline
\smallskip
703 & 053515.4-052040.3 &  31 & 8 (5--11) & 0.07 (0.05--0.14) & 1.5 & 0.70 (5) & 62 &  27.3\,/\,30.0\\
704 & 053515.4-052045.5 &  24 & 0 ($\le$0.3)& 0.33 (0.22--0.45) &
$3\times10^{-5}$ & 1.08 (3) & 36 & 27.1\,/\,27.1 \\
 &  &  & 10 (7--13)&  0.07 ($\le$0.09) & 48 & 1.20 (3) & 31 & 27.1\,/\,30.3 \\
\hline
\end{tabular}
\end{table*}

We used the COUP source catalog and data products of
\citet{getman05b}. A detailed discussion of the COUP data,
encompassing source detection, photon extraction, spectral analysis,
and variability analysis can be found in \citet{getman05b}. We use
{\it HST} archival images to compare the position of COUP sources with
emission knots of HH\,210. 

The best view of the HH\,210 emission system is obtained in the optical
using [\ion{S}{ii}] doublet ($\lambda$6717+$\lambda$6731 lines) or
[\ion{O}{i}] ($\lambda$6300 line), where both bow shocks and a
complicated emission tail are visible, whereas in H$\alpha$ only the
heads of the bow shocks are visible. Figure~\ref{map} shows the
HH\,210 emission system at epoch 2000.70 obtained with the {\it WFPC2}
on board {\it HST} with F673N filter, selecting [\ion{S}{ii}]
\citep[program GO8121NW; ][]{doi02}. The HH\,210 emission system in
the [\ion{O}{i}] line is very similar. We extracted star positions in the
full field-of-view of {\it HST/WFPC2} using {\it Source Extractor}
(version 2.3.2; \citealp{bertin96})\footnote{Available at {\tt
http://terapix.iap.fr}\,.}, and cross-correlated these with COUP
positions. The estimated residual registration error between fifteen
stars in the {\it WPFC2} image and the corresponding COUP sources is
$0\farcs08$. Finally, the total positional accuracy of X-ray sources,
which has a positional uncertainty in COUP of $0\farcs12$, is
$0\farcs14$ in Fig.~\ref{map}. From this {\it WFPC2} image and a
previous one obtained five years before, \citet{doi02} measured the
proper-motion of several [\ion{S}{ii}] knots of HH\,210, ranging from
153\,km\,s$^{-1}$ for 155-040~HH\,210 to 425\,km\,s$^{-1}$ for
154-040a~HH\,210. Using these proper motions we plotted in
Fig.~\ref{map} white arrows pointing at the estimated position of
HH\,210 knots for COUP epoch 2003.04. We conclude that the position of
COUP\,703 is consistent with the position of the [\ion{S}{ii}] knot
154-040a~HH\,210, whereas COUP\,704 is located on the complicated
emission tail of HH\,210 close to an emission filament.

The image of HH\,210 obtained with {\it ACS} on board {\it HST} at
epoch 2004.05 by \citet{bally05} is the closest in time to COUP. This
{\it ACS} observation was made in F658N filter selecting H$\alpha$
($\lambda$6563 line) and [\ion{N}{ii}] ($\lambda$6584 line), and hence
detected mainly the heads of the bow shocks \citep{bally05}. To
register this {\it ACS} image with COUP, we extracted position of {\it
ACS} stars, and cross-correlated these with COUP positions. The
estimated residual registration error between about ninety {\it ACS}
stars and their corresponding COUP sources is $0\farcs12$. Finally,
the total positional accuracy between X-ray sources, which has a
positional uncertainty in COUP of $0\farcs1$, and {\it ACS} sources is
$0\farcs15$. Figure~\ref{trichro} shows the COUP image of HH\,210
using X-ray colors. The contour map indicates the brightest emission
knots observed in H$\alpha$ with {\it HST/ACS}, which were moved
individually to match epoch 2003.04 using proper-motions of
\citet{doi02}. The most southern H$\alpha$ knot has no known
proper-motion; therefore we conservatively assumed for this knot the largest
proper-motion of HH\,210 knots. We conclude that COUP\,703 coincides with
the emission knot with the highest tangential velocity
(425\,km\,s$^{-1}$), 154-040a~HH\,210. In contrast, COUP\,704 is not
associated with any bright H$\alpha$ emission knot, or point source in
the {\it VLT} $J$-, $H$-, $K_{\rm S}$-band survey  (McCaughrean et
al., in preparation) and the {\it IRTF} $L$-band survey
\citep{mccaughrean96}. We note that it
is located near a filament of [\ion{S}{ii}] emission (see Fig.~\ref{map}). 

We compared the spatial distribution of events in COUP\,703 and\,704 with
those in the X-ray counterpart of a nearby star, after adding to this
reference source uniformly distributed background events to match the
background level measured in the extraction regions of COUP\,703
and\,704. Kolmogorov-Smirnov test shows that COUP\,703 and\,704 are
compatible with unresolved sources. Therefore the angular resolution
of \cxo~at $2.7\arcmin$ off-axis, i.e.\, $\sim$1$\arcsec$, is an upper limit
for the size of these objects.

The observed X-ray spectra are shown in Fig.~\ref{spectra}. 
In {\tt XSPEC} (version 11.3), we used
for the spectral fitting between 0.5 and 8\,keV
(corresponding to channels PI=35--548) a photoelectric absorption
({\tt WABS}) combined with a collisionally-ionized plasma model
({\tt APEC}), which provides plasma temperatures down to $0.1$\,MK. 
Table~\ref{fit} gives our best fit parameters. For the source with the
highest net counts, COUP\,703, we found a column density of absorbing
material of $N_{\rm H,21}$$\simeq$8\,cm$^{-2}$ (5--11\,cm$^{-2}$), 
equivalent to $A_{\rm V}$$\sim$5\,mag (3--7\,mag) \citep{vuong03}, and
a plasma temperature of $T$$\simeq$0.8\,MK (0.5--1.6\,MK). For
COUP\,704, the absolute $\chi^2$ minimum corresponds to null absorption and
high plasma temperature, with a low intrinsic X-ray luminosity (see
the right panel of Fig.~\ref{spectra}). This combination of high plasma temperature and
low X-ray luminosity would imply a small radius ($\sim$$10^{14}$\,cm)
of the object driving the shock (see Sect.~\ref{discussion}). However,
another solution with a similar goodness-of-fit exists with higher
absorption and lower plasma temperature, similar to the best-fit
plasma parameters found for COUP\,703. The latter parameters therefore
likely represent a better approximation of the actual physical
conditions in the plasma giving rise to COUP\,704. The
absorption-corrected X-ray luminosities of COUP\,703 and 704 implied
by the spectral fitting are about $10^{30}$\,erg\,s$^{-1}$ (0.5--2.0 keV).

\section{Discussion}
\label{discussion}

HH\,210 displays the largest proper motion among the outflow fingers
extending away from BN-KL, moreover the line emission is blue-shifted,
suggesting a deprojected velocity of about 500\,km\,s$^{-1}$
\citep{axon84,taylor86,hu96}. 154-040a\,HH\,210 is also the brightest
feature in [\ion{O}{iii}] and has a small bow shock attached to it
\citep{odell97,doi02}. The tip of the bow shock is seen as a knot in
[\ion{N}{ii}], H$\alpha$, [\ion{O}{iii}]; hence the tip is the fastest
moving portion of the shock. However, the fact that the [\ion{O}{iii}]
emission requires only a shock with a speed of about 90\,km\,s$^{-1}$
and that the [\ion{O}{iii}] emission is seen only at the tip of the
bow, indicates that this HH object is moving in the wake of other
shocks moving ahead of it \citep{doi02}. 

The observed X-ray emission can be explained with fast moving bow
shocks. The postshock temperature in the adiabatic (non-radiative)
portion of a shock is given by $T_\mathrm{ps}/{\mathrm K}=2.9 \times
10^5/(1+X) \times V^2_{100}$ \citep{Ostriker88}, where $V_{100}$ is
the velocity of the shock front with respect to the downstream
material (the shock velocity) in units of 100\,km\,s$^{-1}$ and $X$ is
the ionization fraction of the preshock gas ($X$=0 for a neutral
medium and 1 for a fully ionized one). Thus, COUP\,703 and
COUP\,704 plasma temperatures of $\sim$0.8\,MK require a shock speed of
$\sim$170\,km\,s$^{-1}$ for a shock moving into a mostly neutral
medium, or $\sim$240\,km\,s$^{-1}$ for a shock moving into a fully
ionized medium. The lower speed is comparable with the bow shock
velocity of 133\,km\,s$^{-1}$ estimated by \citet{odell97} from the
deprojected velocity and the opening angle of the bow wing.

The expected X-ray luminosity of a shock-heated source depends on the
emissivity per unit volume, which depends on the plasma temperature
and density, the volume of the emitting region, and the type of shock,
radiative or non-radiative \citep{raga02}. Neglecting the line
emission, \citet{raga02} find for a radiative shock $L_{\rm r}/{\rm
erg\,s^{-1}}=1.64\times10^{28}\,n_{100}\,r^2_{\rm
b,16}\,V^{5.5}_{100}$~; and for a non-radiative shock $L_{\rm nr}/{\rm
erg\,s^{-1}}=1.8\times10^{29}\,n_{100}^2\,r^3_{\rm b,16}\,V_{100}$,
where $n_{100}$ is the preshock density in units of 100\,cm$^{-3}$ and
$r_{\rm b,16}$ is the radius of the object driving the shock in units
of $10^{16}$\,cm. The X-ray luminosity of a bow shock is the minimum
of these two values. The multiplicative factor to take into account
line emission is 2.5 and 3.0 at 0.1\,MK and 1\,MK, respectively
\citep{raga02}. The {\it HST/ACS} observation of 154-040a~HH\,210
shows a radius of $0\farcs15$ (Fig.~\ref{trichro}), or about
$10^{15}$\,cm at a distance of 450\,pc. Using this dimension for the
X-ray source, a shock speed of 170\,km\,s$^{-1}$, and a preshock
density of about 12000\,cm$^{-3}$, we find $L_{\rm r}$$\sim$$3.3 \times
10^{29}$\,erg\,s$^{-1}$. Thus taking into account line
emission, a bow shock flow can readily explain the X-ray
luminosities observed from COUP\,703 and COUP\,704. 
citet{allen93} found a similar density of $\sim$10$^4$\,cm$^{-3}$
for the electronic density inferred from the [\ion{Fe}{ii}] spectra of
the emission knots.  
The location of HH\,210 inside OMC-1 but close to the limit of the
\ion{H}{ii} region may explain this density and why the medium is not
yet fully ionized by the radiation field of the Trapezium cluster.

We conclude that COUP\,703 is the counterpart of the emission knot
154-040a of HH\,210, and its X-ray emission can be explained by a
radiative bow shock. The X-ray emission of COUP\,704 can
also be explained by a fast-moving, radiative shock toward the tail of
HH\,210. Optical/infrared observations are still needed at other
epochs to reveal in the complicated HH\,210 region the proper motion
of the counterpart to this X-ray source.

\begin{acknowledgements}
We thanks an anonymous referee for useful comments which improved the
paper. COUP is supported by the \cxo{} Guest Observer grant SAO
GO3-4009A (E.\ Feigelson, PI). Further support was provided by the
\cxo{} ACIS Team contract SV4-74018.
\end{acknowledgements}


\end{document}